\documentclass[twocolumn,showpacs,preprintnumbers,amsmath,amssymb]{revtex4}

\usepackage{graphicx}

\begin{document}

\preprint{ULB-TH/05-23,\ hep-th/0512105}

\title{Three dimensional origin of G\"odel spacetimes and black holes}

\author{M\'aximo Ba\~nados, }

\affiliation{Departamento de F\'{\i}sica, P. Universidad Cat\'olica de
Chile, Casilla 306, Santiago 22,Chile}

\author{Glenn Barnich, Geoffrey Comp\`ere}
\affiliation{Physique Th\'eorique et Math\'ematique,
  Universit\'e Libre de Bruxelles and International Solvay
  Institutes, Campus Plaine C.P. 231, B-1050 Bruxelles, Belgium}

\author{Andr\'es Gomberoff} \affiliation{Universidad Nacional
  Andr\'es Bello, Av. Rep\'ublica 239, Santiago, Chile.}

\begin{abstract}

We construct G\"odel-type black hole and particle solutions to
Einstein-Maxwell theory in 2+1 dimensions with a negative cosmological
constant and a Chern-Simons term. On-shell, the electromagnetic
stress-energy tensor effectively replaces the cosmological
constant by minus the square of the topological mass and produces the
stress-energy of a pressure-free perfect fluid. We show how a
particular solution is related to the original G\"odel
universe and analyze the solutions from the point of view of
identifications. Finally, we compute the conserved charges
and work out the thermodynamics.

\end{abstract}


\maketitle

\section{Introduction}

Exact solutions of higher dimensional gravity and supergravity
theories play a key role in the development of string
theory. Recently, a G\"odel-like exact solution of five-dimensional
minimal supergravity having the maximum number of supersymmetries has
been constructed~\cite{Gauntlett:2002nw}. As its four dimensional
predecessor, discovered by G\"odel in 1949~\cite{Godel:1949ga}, this
solution possesses a large number of isometries. It can be lifted to
higher dimensions and has recently been extensively studied as a
background for string and M-theory, see
e.g.~\cite{Boyda:2002ba,Harmark:2003ud}.

The G\"odel-like five-dimensional solution found in
\cite{Gauntlett:2002nw} is supported by an Abelian gauge field. This gauge
field has an additional Chern-Simons interaction and produces the
stress-energy tensor of a pressureless perfect fluid. Since a
Chern-Simons term can also be added in three dimensions, it is a natural
question to ask whether a G\"odel like solution exists in
three-dimensional gravity coupled to a Maxwell-Chern-Simons field.

Actually, there is a stronger motivation to look for this kind
of solutions of three-dimensional gravity. The reason is that the
original four-dimensional G\"odel spacetime is already effectively
three dimensional, see e.g.~\cite{HawkingEllis}. In fact, the metric
has as direct product structure $ds^2_{(4)} = ds_{(3)}^2 + dz^2$ where
$ds_{(3)}^2$ satisfies a purely three-dimensional Einstein equation.

The goal of this paper is twofold. On the one hand we will show that
the three-dimensional factor $ds_{(3)}^2$ of the G\"odel spacetime and
its generalizations \cite{Reboucas:1982hn} are exact solutions of the
three-dimensional Einstein-Maxwell-Chern-Simon theory described by the
action \footnote{It would be interesting to explore in details the
supersymmetric properties of this action and explicitly relate it to
the five dimensional supergravity action.},
\begin{eqnarray}
I = \frac{1}{16\pi G} \int d^3x \left[\sqrt{-g}\left(R
+\frac{2}{l^2} - \frac{1}{4} F_{\mu\nu}F^{\mu\nu}\right)\right. \nonumber\\
\left.-\frac{\alpha}{2}\epsilon^{\mu\nu\rho}A_\mu
F_{\nu\rho}\right]. \label{action}
\end{eqnarray}
The stress-energy tensor of the perfect fluid will be fully generated by
the gauge field $A_\mu$, in complete analogy with the five-dimensional
results reported in \cite{Gauntlett:2002nw}.

Our second goal deals with G\"odel particles and black holes. Within
the five dimensional supergravity theory, rotating black hole
solutions on the G\"odel background have been investigated in
\cite{Herdeiro:2002ft,Gimon:2003ms,Brecher:2003wq,Gimon:2003xk,%
  Behrndt:2004pn,Klemm:2004wq,Gimon:2004if,Barnich:2005kq,Cvetic:2005zi,Konoplya:2005sy}.
It is then natural to ask whether the three-dimensional G\"odel
spacetime $ds_{(3)}^2$ can be generalized to include horizons. This is
indeed the case and a general solution will be displayed
\footnote{M.B.~thanks M.~Henneaux for
  his suggestion to understand these solutions as identifications on
  the G\"odel background.}.

Let us now briefly discuss the general structure of the stress-energy
tensor of Maxwell-Chern-Simons theory. The original G\"odel geometry is a
solution of the Einstein equations in the presence of a pressureless fluid
with energy density $\rho$ and a negative cosmological constant $\Lambda$
such that $\Lambda=-4\pi G\rho$. Equivalently, it can be viewed as a
homogeneous spacetime filled with a stiff fluid, that is,
$p_{SF}=\rho_{SF}=\rho/2$ and vanishing cosmological constant.

In (2+1)-spacetime dimensions, an electromagnetic field can be the
source of such a fluid. To see this it is convenient to write the
stress-energy tensor in terms of the dual field ${}^*\!F^{\mu}$,
\begin{equation}
16\pi G\,T^{\mu\nu} = {}^*\!F^{\mu} {}^*\!F^{\nu} - \frac{1}{2}
{}^*\!F^{\alpha} {}^*\!F_{\alpha}g^{\mu\nu} \ . \label{t}
\end{equation}
In any  region where the field ${}^*\!F^{\mu}$ is timelike, the
electromagnetic field behaves as a
stiff fluid with
\begin{equation}
u^\mu = \frac{1}{\sqrt{-{}^*\!F^{\alpha}
    {}^*\!F_{\alpha}}}{}^*\!F^{\mu},  \ \ \ \
\rho_{SF}=p_{SF}=-{}^*\!F^{\alpha} {}^*\!F_{\alpha}/2.
\label{}
\end{equation}
If G\"odel's geometry is going to be a solution of the
Einstein-Maxwell system, then $\rho_{SF}=-{}^*\!F^{\alpha} {}^*\!F_{\alpha}/2$
must be a constant.
Moreover in comoving
coordinates, in which $g_{tt}=-1$, ${}^*\!F^{\mu}$ must
be a constant vector pointing along the time coordinate.
One can easily
see that such a solution does not exist. In fact, the Maxwell
equations for this solution,
\begin{equation}
d{}^*\!F=0 ,
\label{max}
\end{equation}
imply in these coordinates that $g_{t[\varphi,r]}=0$ which cannot be
achieved for G\"odel. If the electromagnetic field
acquires a topological mass $\alpha$, however, Maxwell's equations
(\ref{max}) will be modified by the addition of the term $\alpha
F$. In that case, the timelike, constant, electromagnetic field is, as
we will see below, a solution of the coupled
Einstein-Maxwell-Chern-Simons system, and the geometry is precisely
that of G\"odel.

Finally, we compute the conserved charges - mass, angular momentum and
electric charge - for these solutions and derive the first laws for
the three dimensional black holes, adapted to an observer at rest with
respect to the electromagnetic fluid. We then show how to adapt this
first law in order to compare with the one for AdS black
holes in the absence of the electromagnetic fluid.

\section{G\"odel spacetime and topologically massive
gravito-electrodynamics }
\label{GODEL}

We start by reviewing the main properties, relevant to our discussion,
of the four-dimensional G\"odel spacetimes
\cite{Godel:1949ga,Reboucas:1982hn,Rooman:1998xf}.
These metrics have a direct product
structure $ds_{(3)}^2+dz^2$ with three-dimensional factor given by
\begin{eqnarray}
ds^2_{(3)} &=& -\left( d t +\frac{4\Omega}{\tilde m^2}\sinh^2{\left(
\frac{\tilde m
   \rho}{2}\right)}d\varphi\right)^2 +d\rho^2 \nonumber \\
 &&\quad + \frac{\sinh^2{(\tilde m \rho)}}{\tilde m^2}d\varphi^2.
\label{Godel}
\end{eqnarray}
The original solution discovered by G\"odel corresponds to
$\tilde m^2=2\Omega^2$. Furthermore, it was pointed out in
\cite{Reboucas:1982hn} that the property of homogeneity and the causal
structure of the G\"odel solution also hold for $\Omega$ and $\tilde m$
independent, provided that $0\leq \tilde m^2 < 4\Omega^2$, the limiting case
$\tilde m^2= 4\Omega^2$ corresponding to anti-de Sitter space.

The three-dimensional metric (\ref{Godel}) has 4 independent Killing
vectors, two obvious ones, $\xi_{(1)}=\partial _t$ and
$\xi_{(2)}=\partial_\varphi$, and two additional ones,
\begin{eqnarray}
 \xi_{(3)} &=& \frac{2\Omega}{\tilde m^2} \tanh(\tilde m\rho/2)
\sin\varphi \frac{\partial}{\partial t}  - \frac{1}{\tilde m} \cos\varphi
\frac{\partial}{\partial \rho} +
\nonumber \\ && \coth(\tilde m\rho) \sin\varphi\frac{\partial}{
\partial \varphi}\,,
 \label{xi3}\\
 \xi_{(4)} &=& \frac{2\Omega}{\tilde m^2} \tanh(\tilde m\rho/2) \cos\varphi
\frac{\partial}{\partial t}
 + \frac{1}{\tilde m} \sin\varphi  \frac{\partial}{\partial \rho} +
 \nonumber\\ && \coth(\tilde m\rho) \cos\varphi \frac{\partial}{
\partial \varphi}.
 \label{xi4}
\end{eqnarray}
which span the algebra $so(2,1) \times \mathbb R$. Finally,
the metric (\ref{Godel}) satisfies the three dimensional Einstein
equations,
\begin{equation}\label{EE}
G^{\mu\nu} - \Omega^2 g^{\mu\nu} = (4\Omega^2 - \tilde
m^2)\delta^\mu_{t} \delta^\nu_t,
\end{equation}
for all values of $\Omega,\tilde m$, and we see that $\Omega$ plays the role
of a negative cosmological constant.

Note that a solution $ds^2_{(3)}$ to Einstein's equations in $3$
dimensions can be lifted to a solution in $4$ dimensions through the
addition of a flat direction $z$ if the additional components of the
stress-energy tensor are chosen as ${\cal T}^{\mu z}=0$ and ${\cal
  T}^{zz}=g_{\mu\nu}{\cal T}^{\mu\nu} +{\Omega^2}/{4\pi G}$. For the
solutions \eqref{Godel}, ${\cal T}^{zz}={(\tilde m^2-2\Omega^2)}/{8\pi G}$
and vanishes, as it should, for the original G\"odel solution.

Our first goal is to prove that (\ref{Godel}) can be regarded as an
exact solution to the equations of motion following from
(\ref{action}).

To this end, we need to supplement (\ref{Godel}) with a suitable gauge
field which will provide the stress-energy tensor (right hand side
of (\ref{EE})). Consider a spherically symmetric gauge field in the
gauge $A_r=0$,
\begin{equation}
A = A_t(\rho) dt  + A_\varphi(\rho) d\varphi.\label{9}
\end{equation}
Inserting this ansatz for the gauge field into the equations of motion
associated to the action (\ref{action}), and assuming that the metric
takes the form (\ref{Godel}), one indeed finds a solution for $A_t$
and $A_\varphi$. Moreover, the two parameters $\tilde m,\Omega$ entering in
(\ref{Godel}) become related to the coupling constants $\alpha$ and
$1/l$ as
\begin{eqnarray}
 \Omega  &=& \alpha\,, \nonumber\\
 \tilde m^2 &=& 2 \left(\alpha^2 + \frac{1}{l^2} \right). \label{trans}
\end{eqnarray}
With this parameterization, the G\"odel sector is determined by
$\alpha^2l^2-1>0$, with $\alpha^2l^2=1$ corresponding to anti-de
Sitter space.  For future convenience, we shall write the solution in
terms of a new radial coordinate $r$ defined by
\begin{equation}
r= {2 \over \tilde m^2}\sinh^2\left( {\tilde m\rho\over 2}\right). \label{r}
\end{equation}
Explicitly, the metric and gauge field are given by,
\begin{eqnarray}
ds^2 &=& -dt^2 - 4\alpha r dt d\varphi + \left[2r-\left(\alpha^2l^2 -
1 \right){2r^2 \over l^2} \right]d\varphi^2 \nonumber \\&&  +
\left(2r+(\alpha^2l^2+1){2r^2\over l^2} \right)^{-1} dr^2\,,  \label{Godel2}\\
A  &=&  \sqrt{\alpha^2l^2 -1}\, {2r \over l}\, d\varphi\,. \label{gauge2}
\end{eqnarray}
{}From now on, we always write $\Omega$ and $\tilde m$ in terms of $\alpha$ and
$l$ using (\ref{trans}).
The general solution for $A$ involves the addition of arbitrary
constant terms along $dt$ and $d\varphi$ in (\ref{gauge2}).
At this stage, we choose the constant in $A_t$ to be zero.  
We will come back to this issue when we discuss
black hole solutions below. A constant term in
$A_\varphi$ is not allowed, however, if one requires $A_\varphi d\varphi$ to be
regular everywhere. Indeed, near $r=0$, the spacelike surfaces of
(\ref{Godel2}) are $\mathbb R^2$ in polar coordinates, the radial
coordinate $r$ in (\ref{Godel2}) being the square root of a standard
radial coordinate over $\mathbb R^2$, and thus $A_\varphi$ must vanish
at $r=0$ because the 1-form $d\varphi$ is not well defined there.

The gauge field (\ref{gauge2}) is also invariant under the isometries
of (\ref{Godel}), up to suitable gauge transformations: for each
Killing vector $\xi^\mu_{(a)}$ there exists a function
$\epsilon_{(a)}$ such that
\begin{equation}\label{} {\cal L}_{\xi_{(a)}} A_\mu
- \partial_\mu \epsilon_{(a)}=0.
\end{equation}
In this sense, the Killing vectors $\xi^\mu_{(a)}$ of (\ref{Godel})
are lifted to gauge parameters $(\xi^\mu_{(a)},\epsilon_{(a)})$ that
leave the full gravity plus gauge field solution invariant. The
generalized G\"odel metric (\ref{Godel2}) together with the gauge
field (\ref{gauge2}) define a background for the action (\ref{action})
with 4 linearly independent symmetries of this type. We shall now use
these symmetries in order to find new solutions describing particles
and black holes (see also \cite{Detournay:2005fz}).

\section{G\"odel particles: $\ {\alpha^2 l^2>1} $}
\label{PARTICLES}

We have proven in the previous section that the G\"odel metric can be
regarded as an exact solution to action (\ref{action}).  The
associated gauge field (\ref{gauge2}) is however real only in the
range $\alpha^2l^2\geq 1$. We consider in this section the case
$\alpha^2l^2>1$ and introduce particle-like objects on the background
(\ref{Godel2}) by means of spacetime identifications.

\subsection{G\"odel Cosmons }

Identifications in three-dimensional gravity were first introduced by
Deser, Jackiw and t'Hooft \cite{Deser:1984tn,Deser:1984dr} and the
resulting objects called ``cosmons". In the presence of
a topologically massive electromagnetic field, cosmons living in a
G\"odel background may also be constructed along these lines.

Take the metric (\ref{Godel2}) and make the following identification
along the Killing vectors $\partial _\varphi$ and $\partial_t$
$$
(t, \varphi) \sim (t- 2  \pi j m,\varphi + 2 \pi m).
$$
where $m,j$ are real constants.
If $m\neq 1$ this procedure will turn
the spatial plane into a cone. The cosmon lives in the tip of this
cone, and its mass is related to $m$ and $j$ (see below). The
time-helical structure given by $j$ will provide angular momentum.

To analyze the resulting geometry it is convenient to pass to a
different set of coordinates,
\begin{eqnarray}
\varphi &=& \varphi' m \nonumber \\
t &=& t' - j\varphi' m  \label{nc}\\
r &=& \frac{r'}{m} +\frac{j}{2\alpha} \nonumber .
\end{eqnarray}
where the above identification amounts to
\begin{equation}\label{}
\varphi' \sim \varphi' + 2\pi n, \ \ \ \ \ n\in Z.
\end{equation}
Also, the new time $t'$ flows ahead smoothly, that is, it does not
jump after encircling the particle.  Inserting these coordinates into
(\ref{Godel2}), and erasing the primes, we find the new metric
\begin{eqnarray}
ds^2 &&= -dt^2 - 4\alpha r dt d\varphi\nonumber\\ && + \left[8G\nu
r-(\alpha^2l^2-1) {2r^2 \over l^2}  - \frac{4GJ}{\alpha} \right]
d\varphi^2 \nonumber \\ &&  +
\left( (\alpha^2l^2+1){2r^2\over l^2} + 8G\nu r - {{4GJ}
\over \alpha}\right)^{-1} dr^2.  \label{particles}
\end{eqnarray}
For fixed  $m$ and $mj$, the new constants $\nu$ and $J$
are given by
\begin{eqnarray}
 4G\nu=m\left(1+\frac{1+\alpha^2l^2}{\alpha l^2}j\right) ,  \label{const1}\\
 4GJ = -m^2 j\left(1+\frac{1+\alpha^2l^2}{2\alpha l^2}j\right).\label{const}
\end{eqnarray}
These constants will be shown to be related to the mass and angular
momentum respectively.

Since under (\ref{nc}) $\varphi$ scales with $m$ while $r$ with
$1/m$ we see that the $r$-dependent part of gauge field
(\ref{gauge2}) is invariant under (\ref{nc}).  However, the
manifold now has a non-trivial cycle, and it is not regular at the
point $r=r_0$ invariant under the action of the Killing vector
whose orbits are used for identifications. Explicitly, $r_0 =
-\frac{jm}{2\alpha}$ which corresponds to $r=0$ before the shift
of $r$ in \eqref{nc}. This means that one can now add a constant
piece to $A_\varphi$. The new gauge field becomes
\begin{equation}\label{}
A = (-\frac{4GQ}{\alpha} + \sqrt{\alpha^2l^2 -1}\, {2r \over l})d\varphi.
\end{equation}
The constant $Q$ will be identified below as the electric charge
of the particle sitting at $r=0$.

The metrics \eqref{particles} only admit the 2 Killing vectors
$\partial_t$ and $\partial_\varphi$. Indeed, the other candidates
$\xi_{(3)}$ and $\xi_{(4)}$ do not survive
as they do not commute with the Killing vector along which the
identifications are made \cite{Banados:1993gq}.

So far we have only used the Killing vectors $\partial /\partial
\varphi$ and $\partial /\partial t$ of (\ref{Godel}) to make
identifications.  Besides these Killing vectors, the metric
(\ref{Godel}) has two other isometries defined by the vectors
(\ref{xi3}) and (\ref{xi4}), and one may consider identifications along
them. We shall not explore this possibility in this paper.

\subsection{Horizons, Singularities and Time Machines}

Distinguished places of the geometry (\ref{ds1}) may appear on those
points where either $g_{\varphi\varphi}$ or $g^{rr}$ vanishes. The
vanishing of $g_{\varphi\varphi}$ indicates that $g_{\varphi\varphi}$
changes sign and hence closed timelike curves (CTC) appear. On the
other hand, the vanishing of $g^{rr}$ indicates the presence of
horizons, as can readily be seen by writting (\ref{particles}) in ADM
form.

The function $g_{\varphi\varphi}$ in (\ref{particles}) is an inverted
parabola, and, it will have two zeros, say $r_1$ and $r_2$ whenever
\begin{equation}
2G\nu^2 > \frac{J(\alpha^2 l^2-1)}{\alpha l^2}.
\label{mu2}
\end{equation}
We must require this condition to be fulfilled in order to have a
``normal" region where $\partial_\varphi$ is spacelike. The boundary of
the normal region are two spacelike surfaces, the velocity of light
surfaces (VLS) at $r=r_1$ and $r=r_2$ (assume $r_2>r_1$). These
surfaces are perfectly regular as long as $g_{t\varphi}\neq 0$
there, which is indeed the case for the metric (\ref{particles}), when
$\alpha\neq 0$.

\begin{figure}
\begin{center}
  \includegraphics[width=8cm]{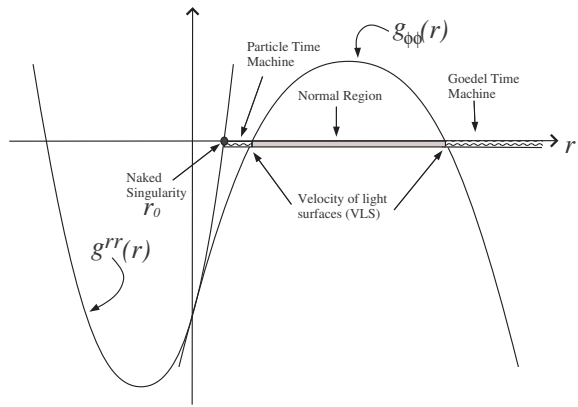}\caption{\label{pneg} G\"odel
    cosmons}
\end{center}
\end{figure}

On the other hand, it is direct to see from (\ref{particles}) that
\begin{equation}\label{grr}
g^{rr}=4\alpha^2 r^2+ g_{\varphi\varphi}.
\end{equation}
Since $g_{\varphi\varphi}$ is positive in the normal region, there are
no horizons there and $g^{rr}$ is positive in that region.  This
together with the fact that $g^{rr}$ is the parabola of
Fig. \ref{pneg}, means that, if any, both zeros of $g^{rr}$ are on the
same side of the normal region. The sides in which no zero of $g^{rr}$
are present are analog to the G\"odel time machine, an unbounded
region, free of singularities, where $\partial_\varphi$ is
timelike. If $\nu\geq 0$, the roots of $g^{rr}$ are smaller than the
roots of $g_{\varphi\varphi}$. Without loss of generality, we can
restrict ourselves to this case because the solutions parametrized by
$(\nu,J,Q)$ are related to those with $(-\nu,J,-Q)$ by the change
of coordinates $r\rightarrow -r$, $\varphi\rightarrow -\varphi$.

The condition for ``would be horizons'' is
\begin{equation}
2G\nu^2 > \frac{J(\alpha^2 l^2+1)}{\alpha l^2}.
\label{mu3}
\end{equation}
As depicted in Fig.~\ref{pneg}, once one reaches the largest root
$r_+=r_0$ of $g^{rr}$, the manifold comes to an end. Indeed, the
signature of the metric changes as one passes $g^{rr}=0$.  This
can be seen by putting the metric in ADM form (see \eqref{ADM}
below). Note that in this case, given $(\nu,J)$, there is a unique
$(m,mj)$ satisfying \eqref{const1}-\eqref{const}.

Using then $r=r_++\kappa_0|\alpha r_+|\rho^2$, with $r_-$ the smallest
root of $g^{rr}$ and
$\kappa_0=\frac{(r_+-r_-)(\alpha^2l^2+1)}{2l^2|\alpha r_+|}$, one
finds near $r_+$,
\begin{eqnarray}
  \label{eq:17}
  ds^2\approx\kappa_0^2
  \rho^2dt^2+d\rho^2-4\alpha^2r_+^2(d\varphi+\frac{dt}{2\alpha
  r_+})^2.
\end{eqnarray}
This means that the spacetime has a naked singularity at
$r_+$, which is the analog of the one found in the
spinning cosmon of \cite{Deser:1984tn,Deser:1984dr}.

Alternatively, as proposed originally in \cite{Cvetic:2005zi} for
the case where the would be horizon is inside the time machine,
one can periodically identify time $t$ with period
$2\pi/\kappa_0$. This leads to
having CTC's lying everywhere, including the normal region.

\section{G\"odel black holes}
\label{BH}

\subsection{The $\alpha^2 l^2 <1 $ sector }

We have shown in Sec.~\ref{GODEL} that the metric (\ref{Godel}) can be
embedded as an exact solution to the equations of motion derived from
(\ref{action}).  The necessary gauge field, given in (\ref{gauge2})
is, however, real only in the range $\alpha^2l^2 \geq 1$.  As we
mentioned in Sec.~\ref{GODEL}, the gauge field (\ref{gauge2})
represents the most general static spherically symmetric solution,
given the metric (\ref{Godel}) (or, in the new radial coordinate,
(\ref{Godel2})).  This means that if we want to find a real gauge
field in the range $\alpha^2 l^2 <1$ we need to start with a different
metric. The goal of this section is to explore the other sector,
$\alpha^2l^2 < 1$, where black holes will be constructed.

Starting from the metric (\ref{Godel2}) and gauge field
(\ref{gauge2}) it is easy to construct a new exact solution which will
be real in the range $\alpha^2 l^2<1$. Consider the following
(complex) coordinate changes \footnote{An equivalent way to do this
  transformation without introducing the imaginary unit is by the
  following sequence of coordinate transformations (and analytic
  continuations) acting on (\ref{Godel2}):
$ t \rightarrow 2t^{1/2}$,  $t \rightarrow -t$,
$ t \rightarrow {1 \over 4}t^2$, and the same for $\varphi$.}
acting on (\ref{Godel2}) and (\ref{gauge2}): $\varphi \rightarrow
i\varphi$,
$t\rightarrow it$, and $r \rightarrow -r$.
The new metric and gauge field read,
\begin{eqnarray}
ds^2 &=& dt^2 - 4\alpha r dt d\varphi + \left[2r-\left(1-\alpha^2l^2
\right) {2r^2\over l^2}\right]
d\varphi^2 \nonumber \\ &&  + \left( (\alpha^2 l^2+1){2r^2\over l^2}
- 2r \right)^{-1} dr^2  \label{Godel3}\\
A &=& \sqrt{1-\alpha^2l^2}\, {2r \over l}d\varphi. \label{gauge3}
\end{eqnarray}

Several comments are in order here. First of all, the intermediate
step of making some coordinates complex is only a way to find a new
solution. From now on, all coordinates $t,r,\varphi$ are defined real,
and, in that sense, the fields (\ref{Godel3}) and (\ref{gauge3})
provide a new exact solution to the action (\ref{action}) which is
real in the range $\alpha^2l^2 <1$.

Second, in the original metric (\ref{Godel2}), the coordinate
$\varphi$ was constrained by the
geometry to have the range $0\leq \varphi < 2\pi$.  This is no longer
the case in the metric (\ref{Godel3}).
The 2-dimensional sub-manifold described by the coordinates
$r,\varphi$ does not have the geometry of $\mathbb R^2$
near  $r\rightarrow 0$ anymore; the coordinate $\varphi$ is thus not
constrained to be compact, and in
principle it should have the full range
\begin{equation}
-\infty < \varphi <  \infty.
\end{equation}
The reason that $\varphi$ in (\ref{Godel3}) is not constrained by the
geometry is that the $g^{rr}$ component of the metric (\ref{Godel3})
changes sign as we approach $r=0$. This is an indication of the
presence of a horizon, although this surface is not yet compact.

Finally, it is worth mentioning that the metrics (\ref{Godel3}) and
(\ref{Godel2}) are real and are related by a coordinate
transformation, so that all local invariants involving the metric
alone have the same values. However, as solutions to the
Einstein-Maxwell equations, they are inequivalent. Indeed, the
diffeomorphism and gauge invariant quantity
$({}^*F)^2={4}(1-\alpha^2l^2)/l^2$ changes sign when going from
(\ref{Godel2})-(\ref{gauge2}) to (\ref{Godel3})-(\ref{gauge3}). This
is different from the pure anti-de Sitter case where particles and
black holes are obtained by identifications performed on the same
background.

\subsection{The G\"odel black
  hole}

Let us go back to (\ref{Godel3}) and note that the function $g^{rr}$
vanishes at $r_+>0$. In order to make the $r=r_+$ surface a regular,
finite area, horizon we shall use the Killing vector $\partial_\varphi$ of
(\ref{Godel3}) to identify points along the $\varphi$ coordinate. In this
case, $\partial _\varphi$ has a non-compact orbit and identifications
along it does not produce a conical singularity, but a ``cylinder".  More
generically, we may proceed in analogy with the cosmon case and identify
along a combination of both $\partial_\varphi$ and $\partial_t $ so that
$$
(t, \varphi) \sim (t- 2  \pi j m,\varphi + 2 \pi m).
$$
so that the resulting geometry will also carry angular momentum.  We
again pass to a different set of coordinates,
\begin{eqnarray}
\varphi &=& \varphi' m \\
t &=& t' - j\varphi' m \\
r &=& \frac{r'}{m} - \frac{j}{2\alpha} ,
\end{eqnarray}
so that the new angular coordinate $\varphi'$ is identified in $2\pi$,
and the time $t'$ flows ahead smoothly.

The new metric reads (after erasing the primes),
\begin{eqnarray}
ds^2 \hspace{-4pt} &=& \hspace{-3pt} dt^2 \hspace{-2pt}- 4\alpha r
dt d\varphi \hspace{-2pt}+\hspace{-2pt} \left( 8G\nu
r-(1-\alpha^2l^2) {2r^2
\over l^2}  - \frac{4GJ}{\alpha} \right) d\varphi^2 \nonumber \\
&& + \left( (\alpha^2l^2+1){2r^2\over l^2} - 8G\nu r +
  \frac{4GJ}{\alpha}\right)^{-1} dr^2.  \label{blackhole}
\end{eqnarray}
As for the particles analyzed in the previous section, for given
$(m,mj)$, we define new
constants $\mu$ and $J$ according to
\begin{eqnarray}
 4G\nu=m\left(1+\frac{1+\alpha^2l^2}{\alpha l^2}j\right) , \\ \label{const5}
 4GJ = m^2 j\left(1+\frac{1+\alpha^2l^2}{2\alpha l^2}j\right).\label{const6}
\end{eqnarray}
Again, these constants will be related below to the mass and the
angular momentum and without loss of generality, we can limit
ourselves to the case $\nu\geq 0$.

In the new coordinates, the electromagnetic potential takes the form
$A=A_\varphi d\varphi$, where
\begin{equation}\label{aphi}
A_\varphi(r) = -\frac{4GQ}{\alpha} + \sqrt{1-\alpha^2l^2}\, {2r \over l}.
\end{equation}
The constant $Q$ is arbitrary because, once again, the nontrivial
topology allows the addition of an arbitrary constant in
$A_\varphi$. It is worth stressing that if $\varphi$ was not compact,
then $m$ and $Q$ would be trivial constants. It also follows that the
Killing vectors of (\ref{Godel3}) have the same form as those of
(\ref{Godel2}), but with the trigonometric functions $\cos(\varphi)$
and $\sin(\varphi)$ replaced by hyperbolic ones. Again, these vectors
do not survive after the identifications.

\subsection{Horizons, Singularities and Time Machines}

\begin{figure}
\begin{center}
\includegraphics[width=9cm]{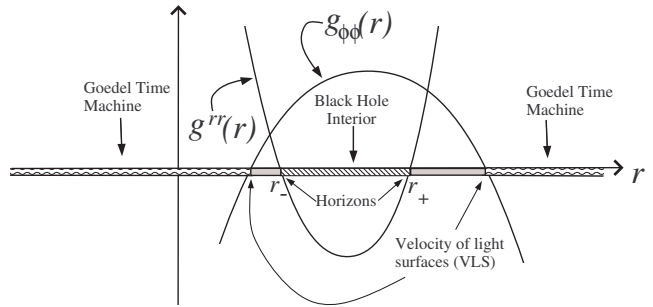}\caption{\label{posp}
G\"odel black holes}
\end{center}
\end{figure}

We now proceed to analyze the metric in the same way we did in the
preceding section.  Again we have a condition for having a normal
region, which, in this case reads
\begin{equation}
2G\nu^2 > \frac{J(1-\alpha^2 l^2)}{\alpha l^2}.
\end{equation}
The functions $g^{rr}$ and $g_{\varphi\varphi}$ now behave as in
Fig.~\ref{posp}.

Note that~\begin{equation}\label{} g^{rr} = - g_{\varphi\varphi}+4\alpha^2
r^2,
\end{equation}
and therefore horizons may only exist in the normal region
of positive $g_{\varphi\varphi}$. Note, however, that for horizons to
exist we must require
\begin{equation}
2G\nu^2 \ge \frac{J(1+\alpha^2l^2)}{\alpha l^2} . \label{reqhor}
\end{equation}
If this requirement is fulfilled, we get two horizons inside the
normal region, $r_-=mj/(2\alpha)$ and $r_+$, which coincide in the
extremal case. The whole normal region is in fact an ergoregion
because $\partial/\partial t$ is spacelike everywhere. Again, for
given $(\nu,J)$, one can then find a unique solution $(m,mj)$ satisfying
\eqref{const5}-\eqref{const6}.

Following Carter \cite{Carter1973}, the metric and the gauge
field can be made regular at both horizons by a combined coordinate and
gauge transformation. Indeed, if $$\Delta(r)=(\alpha^2l^2+1){2r^2\over l^2} -
8G\mu r + \frac{4GJ}{\alpha},$$ the black hole metric can be written as
\begin{eqnarray}
  ds^2 = (dt - 2\alpha r d\varphi)^2 -\Delta d\varphi^2 +
\frac{dr^2}{\Delta}.  \label{blackhole1}
\end{eqnarray}
The analog of ingoing Eddington-Finkelstein coordinates are the
angle $\varphi^{\hspace{-6pt}\leftharpoonup}$ and the time
${t}^{\hspace{-6pt}\leftharpoonup}$ defined by
$d\varphi=d\varphi^{\hspace{-6pt}\leftharpoonup}-\frac{1}{\Delta}dr$,
$dt=dt^{\hspace{-6pt}\leftharpoonup}-\frac{2\alpha r}{\Delta}dr$,
giving the regular metric
\begin{eqnarray}
  ds^2 = (dt^{\hspace{-6pt}\leftharpoonup} - 2\alpha r
  d\varphi^{\hspace{-6pt}\leftharpoonup})^2
  -\Delta d{\varphi^{\hspace{-6pt}\leftharpoonup}}^2
  +2d\varphi^{\hspace{-6pt}\leftharpoonup} dr.  \label{blackhole2}
\end{eqnarray}
With $A_\varphi(r)$ given by \eqref{aphi}, the $r$ dependent gauge
transformation $A^{\hspace{-6pt}\leftharpoonup}=A+d\epsilon$, where
$\epsilon=\int dr \frac{A_\varphi(r)}{\Delta}$ gives the regular
potential $A^{\hspace{-6pt}\leftharpoonup}=A_\varphi(r)d
\varphi^{\hspace{-6pt}\leftharpoonup}$ whose norm
${A^{\hspace{-6pt}\leftharpoonup}}^2$ is zero.

Outgoing Eddington-Finkelstein coordinates are defined by
$d\varphi=-d\varphi^{\hspace{-6pt}\rightharpoonup}+\frac{1}{\Delta}dr$,
$dt=-dt^{\hspace{-6pt}\rightharpoonup}+\frac{2\alpha r}{\Delta}dr$. The
metric then takes also the form~\eqref{blackhole2} with
$t^{\hspace{-6pt}\leftharpoonup}$ and
$\varphi^{\hspace{-6pt}\leftharpoonup}$ replaced by
$-t^{\hspace{-6pt}\rightharpoonup}$ and
$-\varphi^{\hspace{-6pt}\rightharpoonup}$ and the potential can be
regularized by
$A^{\hspace{-6pt}\rightharpoonup}=A-d\epsilon$.

The null generators of the horizons are $\frac{\partial}{\partial
t}+\frac{1}{2\alpha r_\pm}\frac{\partial}{\partial \varphi}$. The
associated ignorable coordinates which are constant on these null
generators are then given by
\begin{equation}
dt^{\pm} = dt - 2\alpha r_{\pm} d\varphi.
\end{equation}
Kruskal type coordinates $(t^{\pm},U^\pm,V^\pm)$ are obtained by defining
\begin{eqnarray}
k_\pm \frac{dV^{\pm}}{V^\pm}
&= & d\varphi^{\hspace{-6pt}\leftharpoonup} = d\varphi + \frac{dr}{\Delta},\\
k_\pm \frac{dU^{\pm}}{U^\pm}& = &
d\varphi^{\hspace{-6pt}\rightharpoonup} = -d\varphi +
\frac{dr}{\Delta },
\end{eqnarray}
where
$$k_\pm = \frac{l^2}{1+\alpha^2l^2}\frac{1}{r_\pm - r_\mp}.$$
In these coordinates, the metric is manifestly regular at the
bifurcation surfaces,
\begin{eqnarray}
ds^2 &= &\big[dt^\pm-\alpha k_\pm (r-r_\mp) (U^\pm
dV^\pm-V^\pm
dU^\pm)\big]^2\nonumber\\
&+&\frac{2k_\pm (r-r_\mp)^2}{r_\pm-r_\mp}dU^\pm dV^\pm,
\end{eqnarray}
with $r$ given implicitly by
\begin{equation}
U^\pm V^\pm = \left( \frac{r-r_+}{r-r_-}\right)^{\pm 1}.
\end{equation}

In Kruskal coordinates, the gauge field \eqref{aphi} becomes
\begin{eqnarray}
  A&=&\frac{k_\pm}{2}
  \left(\frac{A_\varphi(r_\pm)}{U^\pm
  V^\pm}+\frac{\sqrt{1-\alpha^2l^2}}{l}(r-r_\mp) \right)\nonumber\\
  && \times (U^\pm dV^{\pm} -V^\pm dU^\pm).  \label{eq:10}
\end{eqnarray}
The potential can be regularized at $r=r_\pm$ by the
transformations
\begin{eqnarray}
  \label{eq:14}
\hspace*{-8pt}  \tilde A^\pm&=&A-d[A_\varphi(r_\pm)\frac{k_\pm}{2}\ln{
  \frac{V^\pm}{U^\pm}}]\nonumber\\
&=& \frac{k_\pm\sqrt{1-\alpha^2l^2}}{2l}(r-r_\mp)(U^\pm
dV^{\pm}-V^\pm dU^{\pm}).
\end{eqnarray}
In the original coordinates, however, the parameters of these
transformations explicitly involve the angle $\varphi$, $\tilde
A^\pm=A-d[A_\varphi(r_\pm)\varphi]$ and, as explicitly shown below,
they change the electric charge.  
In order to avoid this, one can add a constant piece
proportional to $dt^\pm$, so that
\begin{eqnarray}
  \label{eq:14bis}
  A^\pm&=& \tilde A^\pm - d(\frac{A_\varphi(r_\pm)}{2\alpha
r_\pm}t^\pm).
\end{eqnarray}
In the original coordinates, the gauge parameter is now a linear
function of $t$ alone,
\begin{eqnarray}
A^\pm &=& A - d(\frac{A_\varphi(r_\pm)}{2\alpha
r_\pm}t).\label{time_gauge}
\end{eqnarray}
According to the definition given below, such a transformation does
not change the charges. 

\begin{figure}
\begin{center}
\includegraphics[width=5cm,height=9cm]{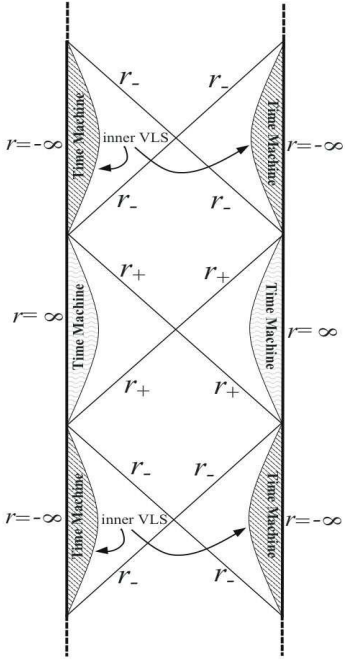}\caption{\label{CP}
Carter-Penrose diagram of G\"odel black hole}
\end{center}
\end{figure}

The causal structure of the G\"odel black hole is displayed in the
Carter-Penrose diagram Fig.~\ref{CP}, where each point represents a
circle.

\section{Vacuum solutions $\alpha^2l^2=1$}

In the case $\alpha^2l^2=1$ the gauge field vanishes and the G\"odel
metric (\ref{Godel2}) reduces to the three-dimensional anti-de Sitter
space (to see this, do the coordinate transformations $\varphi \rightarrow
\varphi+ \alpha t$ and $2r \rightarrow r^2$).  This means that the
identifications in this case yield the usual three-dimensional black
holes, and conical singularities.

\section{The general solution}
\label{section6}

\subsection{Reduced equations of motion}

We have seen in previous sections that the G\"odel metrics
(\ref{Godel}) and (\ref{Godel3}), as well as the corresponding quotient
spaces describing particles and black holes, can be regarded as exact
solutions to the action (\ref{action}).

We have distinguished three cases according to the values of the
dimensionless quantity $\alpha^2l^2$. Our purpose in this section is
to write a general solution which will be valid for all values of
$\alpha^2 l^2$. We shall now construct the solution by looking
directly at the equations of motion.  It is useful to write a general
spherically symmetric static ansatz in the form
\cite{Clement:1993kc,Andrade:2005ur}
\begin{eqnarray}
ds^2 &=& \frac{dr^2}{h^2 - p\, q} + p\, dt^2 + 2h\, dt d\varphi + q\,
d\varphi^2,\label{ds1}
\end{eqnarray}
where $p,q,h$ are functions of $r$ only. This ansatz can also be
written in the ``ADM form",
\begin{equation}\label{ADM}
ds^2   = - \frac{h^2 - pq}{q} dt^2 + \frac{dr^2}{h^2-pq} + q \left(
d\varphi + \frac{h}{q} dt\right)^2.
\end{equation}
This confirms that the function $g^{rr}$
\begin{equation}\label{}
f(r) = h^2(r) - p(r)q(r),
\end{equation}
controls the existence of horizons. Note that for all $p,q,h$, the
determinant of this metric is $\det(-g)=1.$ For the gauge field, we
use the radial gauge $A_r=0$, and assume that $A_t$ and $A_\varphi$
depend only on the radial coordinate,
\begin{equation}
A = A_t(r)\, dt + A_\varphi(r)\, d\varphi.
\end{equation}
In this parametrization, Einstein's equations take the remarkably
simple form,
\begin{eqnarray}
 h'' &=& -A_t'\, A_\varphi' \nonumber \\
 p'' &=& -A_t'^{\, 2} \nonumber\\
 q'' &=& -A_\varphi'^{\, 2} \label{3}\\
 (h^2-pq)''&=&h'^2-p'q' + \frac{4}{l^2},\nonumber
\end{eqnarray}
where primes denote radial derivatives. Maxwell's equations reduce to
\begin{eqnarray}
 (h A_t' - p A_\varphi' - 2\alpha A_t)'=0, \nonumber\\
 (q A_t'-h A_\varphi' - 2\alpha A_\varphi)'=0 . \label{5}
\end{eqnarray}

Before we write the solution to these equations, we make some
general remarks on the structure of the stress-energy tensor
associated to topologically massive electrodynamics. As we pointed out
in the introduction, we will seek for solutions with a constant
electromagnetic field ${}^*\!F$.
Hence, we will only consider potentials $A$ which are linear
in $r$. In this case, Eqs.~(\ref{5}) are
\begin{eqnarray}
 h' A_t' - p' A_\varphi' &=& 2\alpha A_t', \nonumber\\
 q' A_t'-h' A_\varphi' &=& 2\alpha A'_\varphi . \label{dd}
\end{eqnarray}
We now multiply the first by $h'$ and the second by  $p'$, then we subtract
them to obtain
$$
(h'^2-p'q') A_t' = 2\alpha (h'A_t' - p'A'_\varphi) = 4\alpha^2 A_t'.
$$
In the last step we have used Eq.~(\ref{dd}). This implies that, if
$A_t'\neq 0$ then $(h'^2-p'q')=4\alpha^2$. By properly manipulating
Eqs.~(\ref{dd}) we see that this result is also valid if $A_t'=0$ but
$A_\varphi'\neq 0$, and therefore is it true as long as the
electromagnetic field does not vanish.  Now we insert this in the last
equation in (\ref{3}), and obtain,
\begin{eqnarray}
{}^*\!F^\mu {}^*\!F_\mu &=& q(A_t')^2 + p(A_\varphi')^2 -2hA_t'\,
A_\varphi' \nonumber \\ &=& {4\over l^2}\left( 1 -\alpha^2l^2 \right).
\label{f2}
\end{eqnarray}

This equation tells us that when the topological mass $\alpha^2$ is
greater (smaller) than the negative cosmological constant $1/l^2$, the
theory only supports timelike (spacelike) constants fields. Therefore, for
the generalized G\"odel spacetimes (\ref{Godel}), we will need a
topological mass $\alpha^2>1/l^2$. In the other region, the constant
electromagnetic field will describe a tachyonic perfect fluid. Anyway, as
we will see below, it is this region in which black hole solutions are
going to exist.

\subsection{The solution}

By direct computation
one can check that equations (\ref{3})-(\ref{5}) are satisfied by the field
\begin{eqnarray}
p(r)    &=& 8G\mu   \nonumber\\
q(r)   &=& -\frac{4G J}{\alpha}+2r - 2\frac{\gamma^2}{l^2} r^2  \nonumber\\
h(r)   &=& - 2\alpha r   \label{Sol2}\\
A_{t}(r)   &=&  \frac{\alpha^2l^2-1}{\gamma \alpha l}
+ \zeta \nonumber\\
A_{ \varphi}(r)   &=& -\frac{4G}{\alpha} Q + 2\frac{\gamma}{l} r,
\nonumber
\end{eqnarray}
where
\begin{equation}
\gamma = \sqrt{\frac{1-\alpha^2 l^2}{8G\mu}}.
\end{equation}
The parameters $\mu$, $J$ and $Q$ are integration constants with a
physical interpretation as they will be identified with mass, angular
momentum and electric charge below. The arbitrary constant $\zeta$ on
the other hand will be shown to be pure gauge. For later convenience,
it is however useful to keep it along and not restrict ourselves to a
particular gauge at this stage.  This will be discussed in details
Sec.~\ref{conschar}.

In the sector $\alpha^2 l^2 > 1$, the solution is real only for $\mu$
negative. These are the G\"odel particles, i.e., the conical
singularities, discussed in Sec.~\ref{PARTICLES}. The
metric~\eqref{particles} is recovered when $\mu=-2G\nu^2$ and the change
of variables $t \rightarrow t/\sqrt{-8G\mu}$, $r \rightarrow
\sqrt{-8G\mu}\, r$ is performed. For the special values $\mu = -1/8G$ and
$J=0$,  which correspond to the trivial identification $j=0$, $m=1$ in
Sec.~\ref{PARTICLES}, the conical singularities disappear and we are left
with the G\"odel universes~\eqref{Godel2}, used for the identifications
producing the cosmons.

When $\alpha^2 l^2 < 1$, $\mu$ has to be positive. The black hole metrics
\eqref{blackhole} of Sec.~\ref{BH} are recovered when $\mu=2G\nu^2$ and $t
\rightarrow t/\sqrt{8G\mu}$, $r \rightarrow \sqrt{8G\mu}\, r$. For
$\mu=1/8G$ and $J=0$, they reduce to the solution \eqref{Godel3} from
which the black holes have been obtained from non-trivial identifications.

By construction, the electromagnetic stress-energy tensor for the
solutions~\eqref{Sol2} takes the form
\begin{eqnarray}
 \label{eq:TEM}
8\pi G T_{EM}^{\mu\nu}&=& (\alpha^2-\frac{1}{l^2}) g^{\mu\nu}+ 8\pi
G{\mathcal
 T}^{\mu\nu},\\ {\mathcal
 T}^{\mu\nu}&=&\frac{|1-\alpha^2 l^2|}{4\pi G l^2}u^\mu u^\nu,
\label{eq:fluid}
\end{eqnarray}
where the unit tangent vector of the fluid is $u =
\frac{1}{\sqrt{8G|\mu|}}\frac{\partial}{\partial t}$. For $\alpha^2l^2\neq
1$, the effect of the electromagnetic field can be taken into account by
replacing the original cosmological constant $-\frac{1}{l^2}$ by the
effective cosmological constant $-\alpha^2$ and introducing a
pressure-free perfect, ordinary or tachyonic, fluid with energy density
$\frac{|1-\alpha^2l^2|}{4\pi G l^2}$. {}From this point of view, \emph{the
Chern-Simons coupling transmutes into a cosmological constant}. For
$1-\alpha^2l^2 < 0$, the fluid flows along timelike curves while for
$1-\alpha^2l^2 > 0$, the fluid is tachyonic.

When $\alpha^2l^2=1$, the fluid disappears, the stress-energy tensor
vanishes and the solution is real for $\mu \in \mathbb R$.
The metric ~\eqref{Sol2} reduces to the BTZ metric
\cite{Banados:1992wn}, as can be explicitly seen
by transforming to the standard frame that is non-rotating at infinity
with respect to anti-de Sitter space,
\begin{equation}
\varphi \rightarrow \varphi + \alpha t, \qquad r \rightarrow
\frac{r^2}{2}+\frac{2GJ}{\alpha}.\label{frame}
\end{equation}
As will be explained in more details below, in the rotating frame that
we have used, the energy and angular momentum are $\mu$ and $J$
respectively, while they become $M\equiv \mu-\alpha J$ and $J$ in the
standard non-rotating frame.

\begin{figure}
\begin{center}
  \includegraphics[width=6cm]{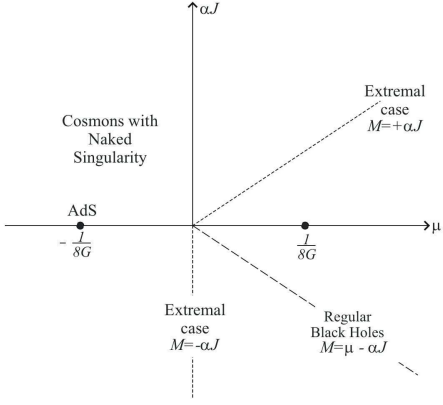}
\caption{\label{diag2} Sectors of the $\alpha^2l^2 = 1$ solution. The BTZ
mass axis $M=\mu-\alpha J$ and the extremal solutions are explicitly
indicated.}
\end{center}
\end{figure}

Regular black holes have the range (see Fig.~\ref{diag2})
 \begin{equation}
 \mu \geq 0,\qquad \mu \geq 2\alpha J.
 \end{equation}
Note that the solution still possess a topological charge $Q$. It has been
discussed in more details in \cite{Andrade:2005ur}.

When $\alpha^2l^2 \neq 1$, the limit $\mu \rightarrow 0$ can be taken
smoothly in the coordinates $\hat r =\gamma r$, $\hat t = t/\gamma$ in
which the solution becomes
\begin{eqnarray}
p(\hat r)    &=& 1-\alpha^2l^2   \nonumber\\
q(\hat r)   &=& -\frac{4G J}{\alpha}+\frac{2}{\gamma}\hat r -
\frac{2}{l^2} \hat r^2  \nonumber\\
h(\hat r)   &=& - 2\alpha \hat r   \label{Sol3}\\
A_{\hat t}(\hat r)   &=&  \frac{\alpha^2l^2-1}{\alpha l}
+ \hat \zeta \nonumber\\
A_{ \varphi}(\hat r)   &=& -\frac{4G}{\alpha} Q + \frac{2}{l} \hat r,
\nonumber
\end{eqnarray}
where $\hat \zeta = \gamma \zeta$.

\begin{figure}
\begin{center}
  \includegraphics[width=6cm]{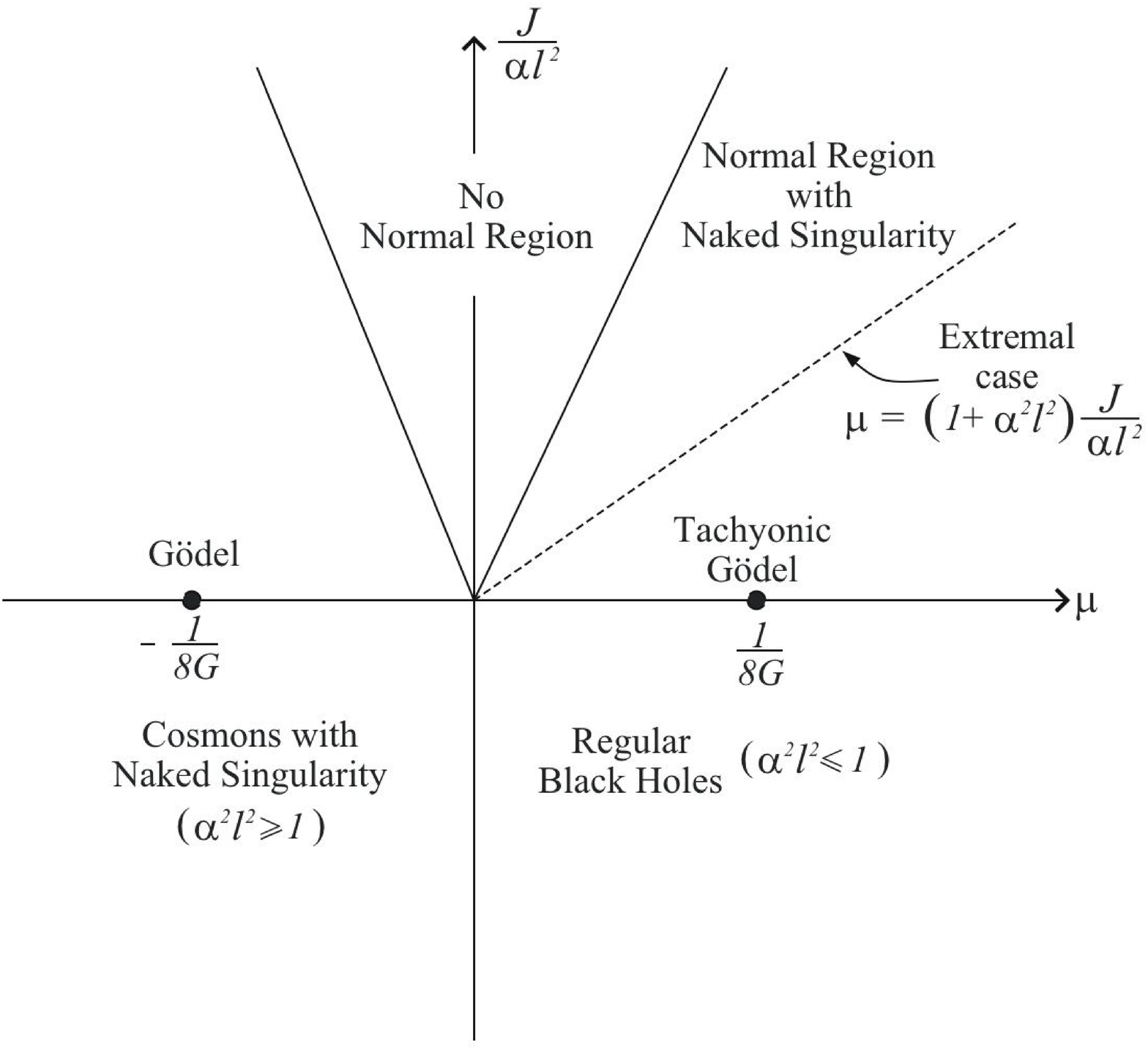}
\caption{\label{diag1} Sectors of the general solution.}
\end{center}
\end{figure}

\section{Conserved charges}
\label{conschar}

\subsection{Angular momentum, electric charge and energies}

The charge differences between a given solution $(g_{\mu\nu}, A_\mu)$ and
an infinitesimally close one $(g_{\mu\nu}+\delta g_{\mu\nu}, A_\mu+\delta
A_\mu)$ are controlled by the linearized theory around
$(g_{\mu\nu},A_\mu)$.  The equivalence classes of conserved $n-2$-forms
(here 1-forms) of the linearized theory can be shown
\cite{Barnich:2001jy,Barnich:2004ts} to be in one-to-one correspondence
with gauge parameters $(\xi^\mu,\epsilon)$ satisfying
\begin{eqnarray}
\left\{\begin{array}{c} \label{eq:2}
 \mathcal{L}_\xi  g_{\mu\nu} = 0,\\
\mathcal{L}_\xi  A_\mu + \partial_\mu
 \epsilon = 0.
\end{array}\right.
\end{eqnarray}

The associated on-shell closed 1-forms can be written as
\cite{Barnich:2005kq} (see also
\cite{Abbott:1981ff,Iyer:1994ys,Anderson:1996sc} for the gravitational
part):
\begin{eqnarray}
 \label{eq:5}
k_{\xi,\epsilon}[(\delta g,\delta A);(g,A)] = k^{grav}_{\xi} + k^{em}_{
 \xi,\epsilon} +  k^{CS}_{\xi,\epsilon},
\end{eqnarray}
where
\begin{eqnarray}
k^{grav}_{ \xi} &=& -\delta K^K_{\xi} + K^K_{\delta
\xi} -\xi\cdot\Theta,\label{eq:6grav}
\end{eqnarray}
where
\begin{eqnarray}
K^K_{\xi}&=& dx^\rho \frac{\sqrt{-g}}{16\pi G } \epsilon_{\rho\mu\nu}
D^\mu\xi^\nu,\label{eq:7}
\end{eqnarray}
is the Komar 1-form and
\begin{eqnarray}
\xi \cdot \Theta&=&dx^\rho \frac{\sqrt{-
 g}}{16\pi G} \epsilon_{\rho\nu\mu}
\xi^{\mu} (g^{\nu\alpha}D^\beta \delta g_{\alpha\beta}-
g^{\alpha\beta}D^\nu \delta g_{\alpha\beta} ). \label{eq:Theta}\nonumber
\end{eqnarray}
The electromagnetic contribution is
\begin{eqnarray}
 k^{em}_{ \xi,\epsilon} = -\delta Q^{em}_{
   \xi,\epsilon}+Q^{em}_{\delta \xi,\delta\epsilon} - \xi \cdot \Theta^{em},
\label{eq:6em}
\end{eqnarray}
where
\begin{eqnarray}
 Q^{em}_{\xi,\epsilon} &=&
 dx^\rho \epsilon_{\rho\mu\nu} \frac{\sqrt{-g}}{32\pi G} \left(
   F^{\mu\nu}(\xi^\rho
   A_\rho +  \epsilon)\right),\label{eq:8bis}\\
\xi \cdot  \Theta^{em}&=&dx^\rho \epsilon_{\rho\nu\mu}\frac{\sqrt{-
     g}}{16\pi G} \xi^\mu  F^{\alpha\nu} \delta A_\alpha .\label{eq:Thetaem}
\end{eqnarray}
The Chern-Simons term contributes as
\begin{equation}
k^{CS}_{ \xi,\epsilon} = dx^\rho \alpha \frac{\sqrt{-g}} {8\pi G} \delta
A_\rho (A_\sigma \xi^\sigma+\epsilon). \label{eq:6CS}
\end{equation}

Finite charge differences are computed by choosing a path $\gamma$ in
parameter space joining the solution $(g,A)$ to a background solution
$(\bar g,\bar A)$ as
\begin{eqnarray}
 \label{eq:4}
 {\mathcal Q}_{\bar\xi,\bar \epsilon} = \oint_S\int_{\gamma}
 k_{\xi,\epsilon}[(\delta g_\gamma,\delta A_\gamma);(g_\gamma,A_\gamma)],
\end{eqnarray}
where $S$ is a closed 1-dimensional submanifold and $(\delta
g_\gamma,\delta A_\gamma)$ denotes the directional derivative of the
fields along $\gamma$ in the space of parameters.  These charges only
depend on the homology class of $S$. They also do not depend on the
path, provided the integrability conditions \cite{Wald:1999wa} $d_V
\oint_S k_{\xi,\epsilon}[(d_V g,d_V A);(g,A)]=0$ are satisfied.

For generic metrics and gauge fields of the form (\ref{Sol2}), the
general solution $(\xi,\epsilon)$ of (\ref{eq:2}) is a linear
combination of $(0,-1)$, $(-\frac{\partial}{\partial \varphi},0)$ and
$(\frac{\partial}{\partial t},0)$.  These basis elements are
associated to infinitesimal charges as follows,
\begin{eqnarray}
 \label{eq:1}
 \oint_S k_{ 0,-1}&=&\delta Q,\
\oint_S k_{-\frac{\partial}{\partial \varphi },0 }=\delta
(J-\frac{2G}{\alpha}Q^2),\
\nonumber\\\oint_S
k_{\frac{\partial}{\partial t},0 }&=&\delta \mu -\zeta\delta Q,
\end{eqnarray}
where the contribution proportional to $\delta Q$ in $\oint_S
k_{-\frac{\partial}{\partial \varphi },0 }$ and $\oint_S
k_{\frac{\partial}{\partial t},0 }$ originate from the Chern-Simons
term through \eqref{eq:6CS}.  The conserved charges associated with
$(0,-1)$, $(-\frac{\partial}{\partial \varphi},0)$ are thus manifestly
integrable.  We choose to associate the angular momentum to
$(-\frac{\partial}{\partial \varphi},-\frac{4GQ}{\alpha})$ so that its
value be algebraically independent of $Q$. If one takes as basis
element $(\frac{\partial}{\partial t},-\zeta)$ instead of
$(\frac{\partial}{\partial t},0)$, one gets a third integrable
conserved charge equal to $\delta \mu$.

The integrated charges computed with respect to the background
$\mu=0=J=Q$ and associated to $(\frac{\partial}{\partial t},-\zeta)$,
$(-\frac{\partial}{\partial
  \varphi},-\frac{4GQ}{\alpha})$ and $(0,-1)$ are the mass, the
angular momentum and the total electric charge respectively,
\begin{eqnarray}
{\mathcal E}=\mu,\ \ {\mathcal
 J}=J,\ {\ \mathcal Q}=Q.
\end{eqnarray}
Note that even though the metric and gauge fields in \eqref{Sol2}
become singular at the background $\mu=0=J=Q$, we can see from the form
\eqref{Sol3}  that this is just a coordinate singularity.

The parameter $\zeta$ is pure gauge because the variation $\delta
\zeta$ is not present in the infinitesimal charges~\eqref{eq:1}.  Note
however that $\zeta$ appears explicitly in the definition of the mass
by associating it with the basis element $(\frac{\partial}{\partial
  t},-\zeta)$. It is only in the gauge $\zeta= 0$, that the mass is
associated with the time-like Killing vector
$(\frac{\partial}{\partial t},0)$. This definition ensures in
particular that the mass of the black hole does not depend on the
gauge transformations (\ref{time_gauge}) needed to regularize the
potential on the bifurcation surfaces.

In order to compare with standard AdS black holes, one has to compute the
mass in the frame \eqref{frame} instead of using the rest frame for the
fluid. The conserved charge $\mathcal E^\prime$ associated with
$(\partial/\partial
t-\alpha
\partial /\partial \varphi,-\zeta+4GQ)$ is now given by
\begin{equation}
\mathcal E^\prime = \mathcal E-\alpha \mathcal J = \mu - \alpha
J=M,
\end{equation}
which coincides with the conventional definition of the mass for the
BTZ black holes.

\subsection{Horizon and first law}
\label{sec:geometry-2}
\subsubsection{General derivation}

\label{sec:general-formulas}

When it exists, the outer horizon $H$ is located at $r_+$, the
largest positive root of $f(r)$. In the following, a subscript
$+$ on a function means that it is evaluated at $r_+$.
The generator of the horizon
is given by $\xi = \frac{\partial}{\partial t} +
\Omega \frac{\partial}{\partial \varphi}$, where the angular
velocity $\Omega$ of the horizon has the value
 \begin{eqnarray}
\Omega=-\varepsilon_{h_+}\varepsilon_{q_+}\sqrt{\frac{p_+}{q_+}}
=-\frac{h_+}{q_+},
\label{eq:9}
\end{eqnarray}
where $\varepsilon_{h_+}$ denotes the sign of $h_+$.
The first law can be derived by starting from
\begin{eqnarray}
\delta{\mathcal E}&=&\oint_{S} k_{\frac{\partial}{\partial
 t},- \zeta} \nonumber\\&=& \oint_{S}k_{ \xi,0}+
 \Omega \oint_{S} k_{-\frac{\partial}{\partial
 \varphi},-\frac{4GQ}{\alpha}}+\oint_{S}k_{- \zeta+\frac{4GQ}{\alpha}\Omega,0}\nonumber\\
&=&\oint_{H}k_{ \xi,0}+  \Omega \delta \mathcal{J} + ( \zeta
-\frac{4GQ}{\alpha}\Omega )\delta {\mathcal Q}.
\end{eqnarray}
The first term on the right-hand side is computed using standard
arguments (see for example \cite{Iyer:1994ys,Gauntlett:1998fz}) to
give
\begin{equation}
\delta \mathcal{E} = \frac{\kappa}{8\pi G} \delta
{\mathcal A} + \Omega
 \delta \mathcal{J}+  \Phi_H^{tot} \delta Q,\label{firstL}
\end{equation}
where the total electric potential is given by
\begin{eqnarray}
 \label{eq:13}
 \Phi_H^{tot} =  \Phi_H +  \zeta
-\frac{4GQ}{\alpha}\Omega,\
 \Phi_H = -( \xi \cdot  A)_+.
\end{eqnarray}
The surface gravity is given by
\begin{eqnarray}
 \kappa=\left.\sqrt{|-\frac{1}{2}(D^\mu \xi^\nu)
     (D_\mu \xi_\nu)|}\right|_H = \frac{|f^\prime_+|}{2\sqrt{|
     q_+|}}, \label{eq:11}
\end{eqnarray}
and the proper area by
\begin{eqnarray}
 \label{eq:12}
\ {\cal A}=2 \pi \sqrt{|q_+|}.\label{s}
\end{eqnarray}
Note that the choice of signs in the definition of electric charge and
angular momentum were made so that the first laws appear in the
conventional form~\eqref{firstL}.

\subsubsection{Explicit values and discussion}
\label{sec:numerical-values}

We have
\begin{eqnarray}
f(r)= 2\frac{(1+\alpha^2 l^2)}{l^2}r^2-16G\mu\left(r -
\frac{2GJ}{\alpha}\right)
\end{eqnarray}
so that
\begin{eqnarray}
 r_+=\frac{4l^2G\mu}{1+\alpha^2l^2}\Big[ 1+\sqrt{1-
\frac{J(1+\alpha^2l^2)}{\alpha l^2\mu}}\Big]\label{rplus} \label{eq:3}
\end{eqnarray}
In order to explicitly verify the first law \eqref{eq:11}, we start by
showing that $\Phi^{tot} = 0$. We need to verify that
\begin{eqnarray}
  \label{eq:6}
 -A_{ t}( r_+)-
 \Omega A_{ \varphi}( r_+)+ \zeta - \Omega\frac{4GQ}{\alpha}=0.
\end{eqnarray}
Using the explicit expressions for the components of $A$, this equation
reduces to
\begin{eqnarray}
  \label{eq:15}
\Omega=\frac{4G\mu}{\alpha r_+}.
 \label{eq:16}
\end{eqnarray}
Taking into account $ \Omega=-{h_+}/{q_+}$ together with $q_+=h^2_+/p_+$,
this equality can then easily be checked using $h_+=-2\alpha r_+$,
$p_+=8G\mu$, implying $q_+=\alpha^2r_+^2/(2G\mu)$. Since
$f^\prime_+=4(1+\alpha^2 l^2)r_+/l^2-16G\mu$, the first law reduces to
\begin{eqnarray}
  \label{eq:16a}
  \delta\mu-\frac{4G\mu}{\alpha r_+}\delta J=
[\frac{\alpha^2l^2+1}{4Gl^2}r_+-\mu][\frac{2\delta
r_+}{r_+}-\frac{\delta\mu}{\mu}],
\end{eqnarray}
which can be explicitly checked using \eqref{rplus}.

In particular, the first law~\eqref{firstL} can be evaluated in the
gauge where the potential is regular on the horizon $r_+$. Because the
two forms (\ref{blackhole}) and (\ref{Sol2}) of the black hole
solution are related by the change of coordinates $t \rightarrow
t\sqrt{-8G\mu}$, $r \rightarrow r/\sqrt{-8G\mu}$, the
gauge~\eqref{time_gauge} now corresponds to
\begin{equation}
A_t =A_t^+= -\Omega A_\varphi^+. \label{value_zeta}
\end{equation}
This amounts to the choice $ \zeta = \frac{4G
Q}{\alpha}\Omega$ in \eqref{Sol2}. It follows that
$\Phi^{tot}=\Phi=0$ and that the vector associated to $A$ is
proportional to $\xi$ on the horizon.

The first law adapted to the energy $\mathcal E^\prime=\mathcal
E-\alpha \mathcal J$
is obtained by
changing $\Omega$ to $\Omega^\prime=\Omega-\alpha$ in
\eqref{firstL}. This form of the first law reduces to the standard
form for 3 dimensional AdS black holes (with or without topological
charge) when $\alpha=\pm 1/l$.

Finally, we note that the first law \eqref{firstL} applies both to the
outer event horizon of a black hole in the normal region and to the
horizon at $r_0$ of a cosmon, when time is identified with real period
$2\pi/|\kappa|$.

\section*{Acknowledgments}


M.~B. wants to thank M.~Henneaux for useful discussions. His work
is supported by Proyectos FONDECYT 1020832 and 7020832. G.~B. and
G.~C. are associated with the National Fund for Scientific
Research, Belgium. Their work is supported in part by a ``P\^{o}le
d'Attraction Interuniversitaire'' (Belgium), by IISN-Belgium,
convention 4.4505.86, by Proyectos FONDECYT 1970151 and 7960001
(Chile) and by the European Commission program
MRTN-CT-2004-005104, in which these authors are associated to
V.U.~Brussels. The work of A.G. is supported in part by FONDECYT
grants 1051084,1051064 and by the research grant 26-05/R of Universidad
Andr\'es Bello.


\begin{thebibliography}{31}
\expandafter\ifx\csname natexlab\endcsname\relax\def\natexlab#1{#1}\fi
\expandafter\ifx\csname bibnamefont\endcsname\relax
  \def\bibnamefont#1{#1}\fi
\expandafter\ifx\csname bibfnamefont\endcsname\relax
  \def\bibfnamefont#1{#1}\fi
\expandafter\ifx\csname citenamefont\endcsname\relax
  \def\citenamefont#1{#1}\fi
\expandafter\ifx\csname url\endcsname\relax
  \def\url#1{\texttt{#1}}\fi
\expandafter\ifx\csname urlprefix\endcsname\relax\def\urlprefix{URL }\fi
\providecommand{\bibinfo}[2]{#2}
\providecommand{\eprint}[2][]{\url{#2}}

\bibitem[{\citenamefont{Gauntlett et~al.}(2003)\citenamefont{Gauntlett,
  Gutowski, Hull, Pakis, and Reall}}]{Gauntlett:2002nw}
\bibinfo{author}{\bibfnamefont{J.~P.} \bibnamefont{Gauntlett}},
  \bibinfo{author}{\bibfnamefont{J.~B.} \bibnamefont{Gutowski}},
  \bibinfo{author}{\bibfnamefont{C.~M.} \bibnamefont{Hull}},
  \bibinfo{author}{\bibfnamefont{S.}~\bibnamefont{Pakis}}, \bibnamefont{and}
  \bibinfo{author}{\bibfnamefont{H.~S.} \bibnamefont{Reall}},
  \bibinfo{journal}{Class. Quant. Grav.} \textbf{\bibinfo{volume}{20}},
  \bibinfo{pages}{4587} (\bibinfo{year}{2003}), \eprint{hep-th/0209114}.

\bibitem[{\citenamefont{G{\"{o}}del}(1949)}]{Godel:1949ga}
\bibinfo{author}{\bibfnamefont{K.}~\bibnamefont{G{\"{o}}del}},
  \bibinfo{journal}{Rev. Mod. Phys.} \textbf{\bibinfo{volume}{21}},
  \bibinfo{pages}{447} (\bibinfo{year}{1949}).

\bibitem[{\citenamefont{Boyda et~al.}(2003)\citenamefont{Boyda, Ganguli,
  Horava, and Varadarajan}}]{Boyda:2002ba}
\bibinfo{author}{\bibfnamefont{E.~K.} \bibnamefont{Boyda}},
  \bibinfo{author}{\bibfnamefont{S.}~\bibnamefont{Ganguli}},
  \bibinfo{author}{\bibfnamefont{P.}~\bibnamefont{Horava}}, \bibnamefont{and}
  \bibinfo{author}{\bibfnamefont{U.}~\bibnamefont{Varadarajan}},
  \bibinfo{journal}{Phys. Rev.} \textbf{\bibinfo{volume}{D67}},
  \bibinfo{pages}{106003} (\bibinfo{year}{2003}), \eprint{hep-th/0212087}.

\bibitem[{\citenamefont{Harmark and Takayanagi}(2003)}]{Harmark:2003ud}
\bibinfo{author}{\bibfnamefont{T.}~\bibnamefont{Harmark}} \bibnamefont{and}
  \bibinfo{author}{\bibfnamefont{T.}~\bibnamefont{Takayanagi}},
  \bibinfo{journal}{Nucl. Phys.} \textbf{\bibinfo{volume}{B662}},
  \bibinfo{pages}{3} (\bibinfo{year}{2003}), \eprint{hep-th/0301206}.

\bibitem[{\citenamefont{Hawking and Ellis}(1973)}]{HawkingEllis}
\bibinfo{author}{\bibfnamefont{S.}~\bibnamefont{Hawking}} \bibnamefont{and}
  \bibinfo{author}{\bibfnamefont{G.}~\bibnamefont{Ellis}},
  \emph{\bibinfo{title}{The large scale structure of space-time}}
  (\bibinfo{publisher}{Cambridge University Press}, \bibinfo{year}{1973}).

\bibitem[{\citenamefont{Rebou{\c{c}}as and Tiomno}(1983)}]{Reboucas:1982hn}
\bibinfo{author}{\bibfnamefont{M.~J.} \bibnamefont{Rebou{\c{c}}as}}
  \bibnamefont{and} \bibinfo{author}{\bibfnamefont{J.}~\bibnamefont{Tiomno}},
  \bibinfo{journal}{Phys. Rev.} \textbf{\bibinfo{volume}{D28}},
  \bibinfo{pages}{1251} (\bibinfo{year}{1983}).

\bibitem[{\citenamefont{Herdeiro}(2003)}]{Herdeiro:2002ft}
\bibinfo{author}{\bibfnamefont{C.~A.~R.} \bibnamefont{Herdeiro}},
  \bibinfo{journal}{Nucl. Phys.} \textbf{\bibinfo{volume}{B665}},
  \bibinfo{pages}{189} (\bibinfo{year}{2003}), \eprint{hep-th/0212002}.

\bibitem[{\citenamefont{Gimon and Hashimoto}(2003)}]{Gimon:2003ms}
\bibinfo{author}{\bibfnamefont{E.~G.} \bibnamefont{Gimon}} \bibnamefont{and}
  \bibinfo{author}{\bibfnamefont{A.}~\bibnamefont{Hashimoto}},
  \bibinfo{journal}{Phys. Rev. Lett.} \textbf{\bibinfo{volume}{91}},
  \bibinfo{pages}{021601} (\bibinfo{year}{2003}), \eprint{hep-th/0304181}.

\bibitem[{\citenamefont{Brecher et~al.}(2003)\citenamefont{Brecher, Danielsson,
  Gregory, and Olsson}}]{Brecher:2003wq}
\bibinfo{author}{\bibfnamefont{D.}~\bibnamefont{Brecher}},
  \bibinfo{author}{\bibfnamefont{U.~H.} \bibnamefont{Danielsson}},
  \bibinfo{author}{\bibfnamefont{J.~P.} \bibnamefont{Gregory}},
  \bibnamefont{and} \bibinfo{author}{\bibfnamefont{M.~E.}
  \bibnamefont{Olsson}}, \bibinfo{journal}{JHEP} \textbf{\bibinfo{volume}{11}},
  \bibinfo{pages}{033} (\bibinfo{year}{2003}), \eprint{hep-th/0309058}.

\bibitem[{\citenamefont{Gimon et~al.}(2003)\citenamefont{Gimon, Hashimoto,
  Hubeny, Lunin, and Rangamani}}]{Gimon:2003xk}
\bibinfo{author}{\bibfnamefont{E.~G.} \bibnamefont{Gimon}},
  \bibinfo{author}{\bibfnamefont{A.}~\bibnamefont{Hashimoto}},
  \bibinfo{author}{\bibfnamefont{V.~E.} \bibnamefont{Hubeny}},
  \bibinfo{author}{\bibfnamefont{O.}~\bibnamefont{Lunin}}, \bibnamefont{and}
  \bibinfo{author}{\bibfnamefont{M.}~\bibnamefont{Rangamani}},
  \bibinfo{journal}{JHEP} \textbf{\bibinfo{volume}{08}}, \bibinfo{pages}{035}
  (\bibinfo{year}{2003}), \eprint{hep-th/0306131}.

\bibitem[{\citenamefont{Behrndt and Klemm}(2004)}]{Behrndt:2004pn}
\bibinfo{author}{\bibfnamefont{K.}~\bibnamefont{Behrndt}} \bibnamefont{and}
  \bibinfo{author}{\bibfnamefont{D.}~\bibnamefont{Klemm}},
  \bibinfo{journal}{Class. Quant. Grav.} \textbf{\bibinfo{volume}{21}},
  \bibinfo{pages}{4107} (\bibinfo{year}{2004}), \eprint{hep-th/0401239}.

\bibitem[{\citenamefont{Klemm and Vanzo}(2005)}]{Klemm:2004wq}
\bibinfo{author}{\bibfnamefont{D.}~\bibnamefont{Klemm}} \bibnamefont{and}
  \bibinfo{author}{\bibfnamefont{L.}~\bibnamefont{Vanzo}},
  \bibinfo{journal}{Fortsch. Phys.} \textbf{\bibinfo{volume}{53}},
  \bibinfo{pages}{919} (\bibinfo{year}{2005}), \eprint{hep-th/0411234}.

\bibitem[{\citenamefont{Gimon and Horava}(2004)}]{Gimon:2004if}
\bibinfo{author}{\bibfnamefont{E.~G.} \bibnamefont{Gimon}} \bibnamefont{and}
  \bibinfo{author}{\bibfnamefont{P.}~\bibnamefont{Horava}}
  (\bibinfo{year}{2004}), \eprint{hep-th/0405019}.

\bibitem[{\citenamefont{Barnich and Comp\`ere}(2005)}]{Barnich:2005kq}
\bibinfo{author}{\bibfnamefont{G.}~\bibnamefont{Barnich}} \bibnamefont{and}
  \bibinfo{author}{\bibfnamefont{G.}~\bibnamefont{Comp\`ere}},
  \bibinfo{journal}{Phys. Rev. Lett.} \textbf{\bibinfo{volume}{95}},
  \bibinfo{pages}{031302} (\bibinfo{year}{2005}), \eprint{hep-th/0501102}.

\bibitem[{\citenamefont{Cvetic et~al.}(2005)\citenamefont{Cvetic, Gibbons, Lu,
  and Pope}}]{Cvetic:2005zi}
\bibinfo{author}{\bibfnamefont{M.}~\bibnamefont{Cvetic}},
  \bibinfo{author}{\bibfnamefont{G.~W.} \bibnamefont{Gibbons}},
  \bibinfo{author}{\bibfnamefont{H.}~\bibnamefont{Lu}}, \bibnamefont{and}
  \bibinfo{author}{\bibfnamefont{C.~N.} \bibnamefont{Pope}},
  \bibinfo{journal}{Phys. Rev. Lett.} \textbf{\bibinfo{volume}{95}},
  \bibinfo{pages}{031302} (\bibinfo{year}{2005}), \eprint{hep-th/0504080}.
  
  \bibitem[{\citenamefont{Konoplya et~al.}(2005)\citenamefont{Konoplya and Abdalla}}]{Konoplya:2005sy}
\bibinfo{author}{\bibfnamefont{R.~A.}~\bibnamefont{Konoplya}},\bibnamefont{and}
  \bibinfo{author}{\bibfnamefont{E.} \bibnamefont{Abdalla}},
  \bibinfo{journal}{Phys. Rev.} \textbf{\bibinfo{volume}{D71}},
  \bibinfo{pages}{084015} (\bibinfo{year}{2005}), \eprint{hep-th/0503029}.

\bibitem[{\citenamefont{Rooman and Spindel}(1998)}]{Rooman:1998xf}
\bibinfo{author}{\bibfnamefont{M.}~\bibnamefont{Rooman}} \bibnamefont{and}
  \bibinfo{author}{\bibfnamefont{P.}~\bibnamefont{Spindel}},
  \bibinfo{journal}{Class. Quant. Grav.} \textbf{\bibinfo{volume}{15}},
  \bibinfo{pages}{3241} (\bibinfo{year}{1998}), \eprint{gr-qc/9804027}.

\bibitem[{\citenamefont{Detournay et~al.}(2005)\citenamefont{Detournay,
  Orlando, Petropoulos, and Spindel}}]{Detournay:2005fz}
\bibinfo{author}{\bibfnamefont{S.}~\bibnamefont{Detournay}},
  \bibinfo{author}{\bibfnamefont{D.}~\bibnamefont{Orlando}},
  \bibinfo{author}{\bibfnamefont{P.~M.} \bibnamefont{Petropoulos}},
  \bibnamefont{and} \bibinfo{author}{\bibfnamefont{P.}~\bibnamefont{Spindel}},
  \bibinfo{journal}{JHEP} \textbf{\bibinfo{volume}{07}},
  \bibinfo{pages}{072} 
  (\bibinfo{year}{2005}), \eprint{hep-th/0504231}.

\bibitem[{\citenamefont{Deser et~al.}(1984)\citenamefont{Deser, Jackiw, and
  't~Hooft}}]{Deser:1984tn}
\bibinfo{author}{\bibfnamefont{S.}~\bibnamefont{Deser}},
  \bibinfo{author}{\bibfnamefont{R.}~\bibnamefont{Jackiw}}, \bibnamefont{and}
  \bibinfo{author}{\bibfnamefont{G.}~\bibnamefont{'t~Hooft}},
  \bibinfo{journal}{Ann. Phys.} \textbf{\bibinfo{volume}{152}},
  \bibinfo{pages}{220} (\bibinfo{year}{1984}).

\bibitem[{\citenamefont{Deser and Jackiw}(1984)}]{Deser:1984dr}
\bibinfo{author}{\bibfnamefont{S.}~\bibnamefont{Deser}} \bibnamefont{and}
  \bibinfo{author}{\bibfnamefont{R.}~\bibnamefont{Jackiw}},
  \bibinfo{journal}{Annals Phys.} \textbf{\bibinfo{volume}{153}},
  \bibinfo{pages}{405} (\bibinfo{year}{1984}).

\bibitem[{\citenamefont{Ba{\~n}ados et~al.}(1993)\citenamefont{Ba{\~n}ados,
  Henneaux, Teitelboim, and Zanelli}}]{Banados:1993gq}
\bibinfo{author}{\bibfnamefont{M.}~\bibnamefont{Ba{\~n}ados}},
  \bibinfo{author}{\bibfnamefont{M.}~\bibnamefont{Henneaux}},
  \bibinfo{author}{\bibfnamefont{C.}~\bibnamefont{Teitelboim}},
  \bibnamefont{and} \bibinfo{author}{\bibfnamefont{J.}~\bibnamefont{Zanelli}},
  \bibinfo{journal}{Phys. Rev.} \textbf{\bibinfo{volume}{D48}},
  \bibinfo{pages}{1506} (\bibinfo{year}{1993}), \eprint{gr-qc/9302012}.

\bibitem[{\citenamefont{Ba\~nados et~al.}(1992)\citenamefont{Ba\~nados,
  Teitelboim, and Zanelli}}]{Banados:1992wn}
\bibinfo{author}{\bibfnamefont{M.}~\bibnamefont{Ba\~nados}},
  \bibinfo{author}{\bibfnamefont{C.}~\bibnamefont{Teitelboim}},
  \bibnamefont{and} \bibinfo{author}{\bibfnamefont{J.}~\bibnamefont{Zanelli}},
  \bibinfo{journal}{Phys. Rev. Lett.} \textbf{\bibinfo{volume}{69}},
  \bibinfo{pages}{1849} (\bibinfo{year}{1992}), \eprint{hep-th/9204099}.

\bibitem[{\citenamefont{Carter}(1973)}]{Carter1973}
\bibinfo{author}{\bibfnamefont{B.}~\bibnamefont{Carter}}, \bibinfo{journal}{in
  Black holes, eds. C. DeWitt and B. S. DeWitt, Gordon and Breach}
  (\bibinfo{year}{1973}).

\bibitem[{\citenamefont{Clement}(1993)}]{Clement:1993kc}
\bibinfo{author}{\bibfnamefont{G.}~\bibnamefont{Clement}},
  \bibinfo{journal}{Class. Quant. Grav.} \textbf{\bibinfo{volume}{10}},
  \bibinfo{pages}{L49} (\bibinfo{year}{1993}).

\bibitem[{\citenamefont{Andrade et~al.}(2005)\citenamefont{Andrade, Ba\~nados,
  Benguria, and Gomberoff}}]{Andrade:2005ur}
\bibinfo{author}{\bibfnamefont{T.}~\bibnamefont{Andrade}},
  \bibinfo{author}{\bibfnamefont{M.}~\bibnamefont{Ba\~nados}},
  \bibinfo{author}{\bibfnamefont{R.}~\bibnamefont{Benguria}}, \bibnamefont{and}
  \bibinfo{author}{\bibfnamefont{A.}~\bibnamefont{Gomberoff}},
  \bibinfo{journal}{Phys. Rev. Lett.} \textbf{\bibinfo{volume}{95}},
  \bibinfo{pages}{021102} (\bibinfo{year}{2005}), \eprint{hep-th/0503095}.

\bibitem[{\citenamefont{Barnich and Brandt}(2002)}]{Barnich:2001jy}
\bibinfo{author}{\bibfnamefont{G.}~\bibnamefont{Barnich}} \bibnamefont{and}
  \bibinfo{author}{\bibfnamefont{F.}~\bibnamefont{Brandt}},
  \bibinfo{journal}{Nucl. Phys.} \textbf{\bibinfo{volume}{B633}},
  \bibinfo{pages}{3} (\bibinfo{year}{2002}),
  \eprint[http://arXiv.org/abs]{hep-th/0111246}.

\bibitem[{\citenamefont{Barnich et~al.}(2004)\citenamefont{Barnich, Leclercq,
  and Spindel}}]{Barnich:2004ts}
\bibinfo{author}{\bibfnamefont{G.}~\bibnamefont{Barnich}},
  \bibinfo{author}{\bibfnamefont{S.}~\bibnamefont{Leclercq}}, \bibnamefont{and}
  \bibinfo{author}{\bibfnamefont{P.}~\bibnamefont{Spindel}},
  \bibinfo{journal}{Lett. Math. Phys.} \textbf{\bibinfo{volume}{68}},
  \bibinfo{pages}{175} (\bibinfo{year}{2004}), \eprint{gr-qc/0404006}.

\bibitem[{\citenamefont{Abbott and Deser}(1982)}]{Abbott:1981ff}
\bibinfo{author}{\bibfnamefont{L.~F.} \bibnamefont{Abbott}} \bibnamefont{and}
  \bibinfo{author}{\bibfnamefont{S.}~\bibnamefont{Deser}},
  \bibinfo{journal}{Nucl. Phys.} \textbf{\bibinfo{volume}{B195}},
  \bibinfo{pages}{76} (\bibinfo{year}{1982}).

\bibitem[{\citenamefont{Iyer and Wald}(1994)}]{Iyer:1994ys}
\bibinfo{author}{\bibfnamefont{V.}~\bibnamefont{Iyer}} \bibnamefont{and}
  \bibinfo{author}{\bibfnamefont{R.~M.} \bibnamefont{Wald}},
  \bibinfo{journal}{Phys. Rev.} \textbf{\bibinfo{volume}{D50}},
  \bibinfo{pages}{846} (\bibinfo{year}{1994}), \eprint{gr-qc/9403028}.

\bibitem[{\citenamefont{Anderson and Torre}(1996)}]{Anderson:1996sc}
\bibinfo{author}{\bibfnamefont{I.~M.} \bibnamefont{Anderson}} \bibnamefont{and}
  \bibinfo{author}{\bibfnamefont{C.~G.} \bibnamefont{Torre}},
  \bibinfo{journal}{Phys. Rev. Lett.} \textbf{\bibinfo{volume}{77}},
  \bibinfo{pages}{4109} (\bibinfo{year}{1996}), \eprint{hep-th/9608008}.

\bibitem[{\citenamefont{Wald and Zoupas}(2000)}]{Wald:1999wa}
\bibinfo{author}{\bibfnamefont{R.~M.} \bibnamefont{Wald}} \bibnamefont{and}
  \bibinfo{author}{\bibfnamefont{A.}~\bibnamefont{Zoupas}},
  \bibinfo{journal}{Phys. Rev.} \textbf{\bibinfo{volume}{D61}},
  \bibinfo{pages}{084027} (\bibinfo{year}{2000}), \eprint{gr-qc/9911095}.

\bibitem[{\citenamefont{Gauntlett et~al.}(1999)\citenamefont{Gauntlett, Myers,
  and Townsend}}]{Gauntlett:1998fz}
\bibinfo{author}{\bibfnamefont{J.~P.} \bibnamefont{Gauntlett}},
  \bibinfo{author}{\bibfnamefont{R.~C.} \bibnamefont{Myers}}, \bibnamefont{and}
  \bibinfo{author}{\bibfnamefont{P.~K.} \bibnamefont{Townsend}},
  \bibinfo{journal}{Class. Quant. Grav.} \textbf{\bibinfo{volume}{16}},
  \bibinfo{pages}{1} (\bibinfo{year}{1999}), \eprint{hep-th/9810204}.

\end{thebibliography}


\end{document}